\documentclass[preprint,aps,pra,superscriptaddress]{revtex4-1}

\usepackage{array}
\usepackage{relsize}
\usepackage{placeins}
\usepackage{amsmath}
\usepackage{mathrsfs}
\usepackage{amssymb}
\usepackage{graphicx}
 \usepackage{color}
  \usepackage{hyperref}
  \usepackage{mathtools}

\usepackage[normalem]{ulem}

\usepackage[font=small, labelfont=sc]{caption}
\usepackage[font=small, labelfont=sc]{subcaption}
\usepackage{dcolumn}
 \usepackage{color}

\newlength{\sfigsize}
\newlength{\lfigsize}
\newcommand{\be}{\begin{equation}}
\newcommand{\ee}{\end{equation}} 
\newcommand{\lb}{\label}
\newcommand{\OL}{\overline}
\newcommand{\wt}{\widetilde}
\newcommand{\wh}{\widehat}

\newcommand{\mModel}{\mbox{\small m}}

\newcommand{\const}{({\rm const.})}

\newcommand{\ba}{{\bf a}}

\newcommand{\bF}{{\bf F}}

\newcommand{\br}{{\bf r}}

\newcommand{\bu}{{\bf u}}

\newcommand{\bx}{{\bf x}}

\newcommand{\bJ}{{\bf J}}
\newcommand{\bS}{{\bf S}}

\newcommand{\bomega}{{\mbox{\boldmath $\omega$}}}

\newcommand{\grad}{{\mbox{\boldmath $\nabla$}}}
\newcommand{\bdot}{{\mbox{\boldmath $\cdot$}}}

\newcommand{\btimes}{{\mbox{\boldmath $\times$}}}

\begin{document}
\title{\textbf {Baropycnal Work: A Mechanism for Energy Transfer Across Scales}}
\author{Aarne Lees}
\affiliation{Department of Mechanical Engineering, University of Rochester}
\affiliation{Laboratory for Laser Energetics, University of Rochester}
\author{Hussein Aluie}
\affiliation{Department of Mechanical Engineering, University of Rochester}
\affiliation{Laboratory for Laser Energetics, University of Rochester}
\email{hussein@rochester.edu}




\begin{abstract}
The role of baroclinicity, which arises from the misalignment of pressure and density gradients, is well-known in the vorticity equation, yet its role in the kinetic energy budget has never been obvious. Here, we show that baroclinicity appears naturally in the kinetic energy budget after carrying out the appropriate scale decomposition. Strain generation by pressure and density gradients, both barotropic and baroclinic, also results from our analysis. These two processes underlie the recently identified mechanism of ``baropycnal work,'' which can transfer energy across scales in variable density flows. As such, baropycnal work is markedly distinct from pressure-dilatation into which the former is implicitly lumped in Large Eddy Simulations. We provide numerical evidence from $1{,}024^3$ direct numerical simulations of compressible turbulence. The data shows excellent pointwise agreement between baropycnal work and the nonlinear model we derive, supporting our interpretation of how it operates.
\end{abstract}

\maketitle

 \section{Introduction}

Energy transfer across length scales is one of the defining characteristics of turbulent flows, the subject of which fits well
under the ``Multiscale Turbulent Transport'' theme of this special issue in \emph{Fluids}. 
In constant density turbulence, the only pathway for transferring energy across scales is \emph{deformation work} \cite{TennekesLumley72,AlexakisBiferale18}, which we represent below by $\Pi$. This is often referred to as the \emph{turbulence production} term in the turbulent kinetic energy (TKE) budget within the Reynolds averaging decomposition \cite{Pope00} or the \emph{spectral flux} within the Fourier decomposition \cite{Frisch95,Alexakisetal05b,Mininni06}. Deformation work gives rise to the cascade in incompressible turbulence, which is largely believed to operate by vortex stretching in 3-dimensions \cite{TennekesLumley72,BorueOrszag98,Eyink06b,Eyink08,Xuetal11}, an idea which may be traced back to G.I Taylor \cite{Taylor37,Taylor38}. 

Recent studies \cite{Aluie11,Aluieetal12,Aluie13,Kritsuketal13,Wangetal13,EyinkDrivas18} have shown that in the presence of density variations, such as in compressible turbulence, there exists another pathway across scales called ``baropycnal work,'' represented by $\Lambda$ below. In the traditional formulation of the compressible Large Eddy Simulation (LES) equations \cite{Favre69,Lele94,Garnieretal09,OBrienetal2014}, which are essentially a coarse-graining decomposition of scales, $\Lambda$ is almost always implicitly lumped with $P\grad\bdot\bu$, where $P$ is pressure and $\bu$ is velocity, and treated as a large scale (resolved) pressure-dilatation which does not require modeling. Ref. \cite{Aluie13} argued that baropycnal work, $\Lambda$, is more similar in nature to deformation work, $\Pi$, in that it involves large-scales interacting with small-scales thereby allowing it to transfer energy \emph{across} scales. As such, it is fundamentally distinct from pressure dilatation which involves only large-scales and cannot transfer energy directly across scales.

In this work, we shall investigate the mechanisms by which baropycnal work transfers energy across scales. The main result is embodied in eq. (\ref{eq:lambda_decomp}) below, which shows that $\Lambda$ transfers energy by two processes: 
\begin{enumerate}
\item Barotropic and baroclinic generation of strain, $\bS$, from gradients of pressure and density, $\rho$:\\
$\const\,\ell^2\,\rho^{-1}\left[\grad P \bdot\bS\bdot\grad\rho\right] = \const\,\ell^2\,\rho^{-1}\left[\left(\grad\rho\left(\grad P\right)^{T}\right){\bf :} \bS\right] $,
\item Baroclinic generation of vorticity, $\bomega$:\\ 
$\const\,\ell^2\,\rho^{-1}\left(\grad\rho\btimes\grad P\right)\bdot \bomega$, 
\end{enumerate}
where the dyadic product $\grad\rho\left(\grad P\right)^{T}$ is a tensor. Length scale $\ell$ is that  at which density and pressure gradients are evaluated as we make clear in eqs. \eqref{eq:lambda_decomp}-\eqref{eq:Lambda_BC} below.
To our knowledge, these results are the first to show how baroclinicity, $\grad\rho\btimes\grad P$, appears in the kinetic energy budget. Baroclinicity is often analyzed within the vorticity budget but its role in the kinetic energy budget has never been obvious. We shall show here that baroclinicity appears naturally in the kinetic energy budget after performing the appropriate scale decomposition. Strain generation by pressure and density gradients (both barotropic and baroclinic) also results from our analysis, highlighting its potential significance which is often overlooked in the literature.
The processes of strain and vortex generation show baropycnal work $\Lambda$ to be markedly distinct from the process of pressure dilatation, further supporting the argument against lumping the two terms. 

There is a diverse array of applications in this research subject. Density variability can arise in high Mach number flows, 
but it is also pertinent in the limit of low Mach numbers along contact discontinuities such as in
multi-species or multi-phase flows \cite{Mukherjeeetal18}. Variable density (VD) flows 
are relevant in a wide range of systems, such as 
in molecular clouds in the interstellar medium \cite{Kritsuketal07,Federrathetal10,Panetal16}), in inertial confinement fusion \cite{Yanetal16, Zhangetal18a,Zhangetal18b}, in high-speed flight and combustion \cite{Larssonetal15,Urzay18},
and in air-sea interaction in geophysical flows \cite{Emanuel86,Bourassaetal16,Deikeetal16,Renaultetal17}.

The outline of the paper is as follows.
In section \ref{sec:Filtering}, we shall use coarse-graining to decompose scales and identify baropycnal work, $\Lambda$.
In section \ref{sec:Mechanism}, we use scale locality to approximate $\Lambda$ with a nonlinear model, which in turn shows 
how $\Lambda$ is due to a combination of strain generation and baroclinic vorticity generation.
In section \ref{sec:simulations}, we describe our direct numerical simulations (DNS) used in section \ref{sec:NumericalResults}  to present evidence that indeed $\Lambda$ and its nonlinear model exhibit excellent agreement. This justifies our analysis of $\Lambda$ via its nonlinear model, similar to what was done by \cite{BorueOrszag98} in their analysis of $\Pi$. We conclude with section \ref{sec:Conclusion}.

\section{Multi-Scale Dynamics}\lb{sec:Filtering}
To analyze the dynamics of different scales in a compressible flow, we use the coarse-graining approach,
which has proven to be a natural and versatile framework to understand and model scale interactions 
(e.g. \cite{MeneveauKatz00,Eyink05}). 
The approach is standard in partial differential equations
and distribution theory (e.g., see Refs.\cite{Strichartz03,Evans10}). 
It became common in Large Eddy Simulation (LES) modeling of turbulence
thanks to the original work of Leonard \cite{Leonard74} and the later work of
Germano \cite{Germano92}.
Eyink \cite{Eyink95a,Eyink95b,Eyink05,EyinkAluie09} 
subsequently developed the formalism mathematically to analyze the fundamental physics of 
scale coupling in turbulence.

Coarse-graining has been used in many fluid dynamics 
applications, ranging from DNS of incompressible turbulence 
 \citep[e.g.][]{Piomellietal91,Vremanetal94,AluieEyink09,Buzzicottietal18a}, to 
 2D laboratory flows \citep[e.g.][]{Riveraetal03,Chenetal03,Chenetal06,KelleyOuellette11,Riveraetal14,LiaoOuellette15,FangOuellette16},
 to experiments of turbulent jets \citep{Liuetal94} and flows through a grid \citep{Meneveau94}, through a duct \citep{Taoetal02}, in a water channel \citep{Baietal13}, and in turbomachinery \citep[e.g.][]{Chowetal05,AkbariMontazerin13}. Moreover, the framework has been extended to geophysical flows \citep{AluieKurien11,Aluieetal18,Buzzicottietal18b}, magnetohydrodynamics \citep{AluieEyink10,Aluie17}, and compressible turbulence \citep[e.g.][]{Aluieetal12}, and most recently as a framework to extract the spectrum in a flow \cite{SadekAluie18}.

For any field $\ba(\bx)$, a coarse-grained or (low-pass) filtered field, which contains modes
at scales $>\ell$, is defined in $n$-dimensions as
\be
\OL \ba_\ell(\bx) = \int d^n\br~ G_\ell(\br)\, \ba(\bx+\br),
\lb{eq:filtering}\ee
where $G(\br)$ is a normalized convolution kernel and $G_\ell(\br)= \ell^{-n} G(\br/\ell)$ is a dilated version of the kernel having its main support over a region of diameter $\ell$. The scale decomposition in (\ref{eq:filtering}) is essentially a partitioning of scales in the system into large ($\gtrsim\ell$), captured by $\OL \ba_\ell$, and small ($\lesssim\ell$), captured by the residual $\ba'_\ell=\ba-\OL \ba_\ell$. In the remainder of this paper, we shall omit subscript $\ell$ from variables if there is no risk for confusion.

\subsection{Variable Density Flows}
In incompressible turbulence, our understanding of the scale dynamics of kinetic energy centers on analyzing $|\OL{\bu}_\ell|^2/2$. 
In variable density turbulence, scale decomposition is not as straightforward due to the density field $\rho(\bx)$.
Several definitions of ``large-scale'' kinetic energy have been used in the literature, corresponding to different
scale-decompositions as discussed in \cite{Aluie13}. These include $\OL\rho_\ell |\OL\bu_\ell|^2/2$, 
which has been used in several studies (e.g. \cite{chassaing1985alternative,BodonyLele05,Burton11,KarimiGirimaji17}), and
$|\OL{(\sqrt{\rho}\bu)}_\ell|^2/2$, which has also been used extensively in compressible turbulence studies
(e.g. \cite{kida1990energy,CookZhou02,Wangetal13,Greteetal17}). A ``length-scale'' within these different 
decompositions corresponds to different flow variables, each of which can yield quantities with units of energy. 
However, as demonstrated by \cite{ZhaoAluie18}, such decompositions can violate the so-called \emph{inviscid criterion}, yielding difficulties with disentangling viscous from inertial dynamics in turbulent flows. The {inviscid criterion} stipulates that a scale decomposition should guarantee a negligible contribution from viscous terms in the evolution equation of the large length-scales.
It was shown mathematically in \cite{Aluie13} and demonstrated numerically in \cite{ZhaoAluie18} that 
a Hesselberg-Favre decomposition, introduced by Hesselberg \cite{Hesselberg26} but often associated with Favre \cite{favre1958further,Favre69}, $|\OL{\rho\bu}_\ell|^2/2\OL{\rho}_\ell$,  satisfies the inviscid criterion, which allows for properly disentangling the dynamical ranges of scales. 
We will use the common notation
\be
\wt\bu_\ell(\bx) = \OL{\rho \bu}_\ell/\OL\rho_\ell~~,
\lb{eq:Favrefiltering}\ee
which yields $\OL\rho_\ell|\wt{\bu}_\ell|^2/2$ for kinetic energy at scales larger than $\ell$. 
The budget for the large-scale KE can be easily derived \cite{Aluie13} from the momentum equation \eqref{momentum} below:
\be
\partial_t \OL\rho_\ell\frac{|\wt\bu_\ell|^2}{2} + \grad\bdot\bJ_\ell
= -\Pi_\ell -\Lambda_\ell + \OL{P}_\ell\grad\bdot\OL\bu_\ell
-D_\ell
+\epsilon^{inj}_\ell,
\lb{largeKE}\ee
where $\bJ_\ell(\bx)$ is space transport of large-scale kinetic energy, $-\OL{P}_\ell\grad\bdot\OL\bu_\ell$ is large-scale
pressure dilatation, $D_\ell(\bx)$ is viscous dissipation acting
on scales $>\ell$, and $\epsilon^{inj}_\ell(\bx)$ is the energy injected due to external stirring. These terms are defined in eqs. (16)-(18) of Ref. \cite{Aluie13}. The $\Pi_\ell(\bx)$ and $\Lambda_\ell(\bx)$ terms account for the transfer of energy \emph{across} scale $\ell$, and are defined as 
\begin{eqnarray}
&\Pi_\ell(\bx)& = ~  -\OL{\rho}~ \partial_j\wt{u}_i  ~\wt\tau(u_i,u_j) ~~ \lb{eq:DeformationWork}\\
& \Lambda_\ell(\bx)& = ~  \frac{1}{\OL\rho}\partial_j\OL{P}~\OL\tau(\rho,u_j), ~~ \lb{eq:BaropycnalWork}
 \end{eqnarray}
where \be  \OL\tau_\ell(f,g) \equiv\OL{(fg)}_\ell-\OL{f}_\ell\OL{g}_\ell \lb{tau-def} \ee
is a $2^{nd}$-order \emph{generalized central moment} of fields $f(\bx),g(\bx)$ (see \cite{Germano92}).

The first flux term, $\Pi_\ell$, is similar to its incompressible counterpart and is often called deformation work. The second flux term, $\Lambda_\ell$, was identified in \cite{Aluie11,Aluie13} and called ``baropycnal work.'' It is inherently due to the presence of a variable density and 
vanishes in the incompressible limit.
Recent work by Eyink and Drivas \cite{EyinkDrivas18,DrivasEyink18} identified a third possible pathway for energy transfer, which they called 
``pressure-dilatation defect'' and arises when the joint limits of $\kappa,\mu\to 0$ and $\ell\to0$ do not commute. Eyink and Drivas \cite{EyinkDrivas18} showed that the pressure-dilatation defect mechanism transfers energy downscale in 1D normal shocks. In this paper, we shall focus on understanding the mechanisms by which baropycnal work transfers energy across scales.

\section{The Mechanism of Baropycnal Work}\lb{sec:Mechanism}
Our investigation of the mechanism behind baropycnal work is inspired by the work of Borue \& Orszag
 \cite{BorueOrszag98}, where they used the nonlinear model of the energy flux $\Pi_\ell$ to show 
that it operates, on average, by vortex stretching (see also eq. (32) in \cite{Eyink06b}). 

Our derivation of the nonlinear model of $\Lambda_\ell$ will follow that in \cite{Eyink06a}, 
which is somewhat different from the standard derivation of nonlinear models \cite{Bardinaetal80,Liuetal94,BorueOrszag98}.
We utilize the property of scale-locality \cite{Eyink05} (specifically, ultraviolet locality) of the subscale mass flux which was proved to hold in variable density flows by \cite{Aluie11} under weak assumptions. Specifically, for our present purposes, we require that the spectra of density and velocity decay faster than $k^{-1}$ in wavenumber. In other words, density and velocity should have finite second-order moments, $\langle\rho^2\rangle<\infty$ and $\langle|\bu|^2\rangle<\infty$, in the limit of infinite Reynolds number. Here, $\langle\dots\rangle$ is a space average.
Ultraviolet scale locality implies that contributions to the subscale mass flux $\overline{\tau}_{\ell}(\rho, \bu)$ at scale $\ell$ from smaller scales $\delta\ll \ell$ are negligible \cite{Aluie11}:
\be
|\overline{\tau}_{\ell}(\rho'_\delta, \bu'_\delta)| \ll |\overline{\tau}_{\ell}(\rho, \bu)|
\lb{eq:UVlocality}\ee
By assuming the validity of eq.\eqref{eq:UVlocality} in the limit $\delta\to\ell$, we can justify the approximation
\begin{equation}
  \overline{\tau}_{\ell} (\rho, \bu) \approx \overline{\tau}_{\ell} \left (
    \overline{\rho}_{\ell}, \overline{\bu}_{\ell} \right ),
\end{equation}
which neglects any contribution from scales $<\ell$ to the subscale mass flux.
Using the usual definition of an increment:
\be\delta f(\bx;\br) = f (\mathbf{x}+\mathbf{r}) - f (\mathbf{x}),
\ee
the subscale mass flux term can be rewritten exactly in terms of $\delta \rho$ and $\delta \bu$ \cite{Eyink05,Aluie11}:
\begin{equation}
  \overline{\tau} (\overline{\rho}_{\ell}, \overline{\bu}_{\ell})
    = \left\langle \delta \overline{\rho}_{\ell} \, \delta
      \overline{\bu}_{\ell} \right\rangle_{\ell}
    - \left\langle\delta \overline{\rho}_{\ell} \right\rangle_{\ell}
      \left\langle \delta \overline{\bu}_{\ell} \right\rangle_{\ell}
\lb{eq:SubgridTauDeltas}\end{equation}
Equation \eqref{eq:SubgridTauDeltas} is exact, where 
\be
\left\langle \delta \overline{f}_{\ell}(\bx;\br) \right\rangle_{\ell} = \int d\br\, G_\ell(\br) \delta \overline{f}_{\ell}(\bx;\br)
\ee
 is a local average around $\bx$ over all separations $\br$ weighted by the kernel $G_\ell$. A spatially localized kernel effectively limits the average to separations $|\br|\lesssim\ell/2$.
 Since a filtered field $\OL{f}_\ell(\bx)$ is smooth, we can 
Taylor expand its increments around $\bx$ 
\begin{equation}
  \delta \OL{f}(\bx;\br) = \OL{f}(\bx+\br)-\OL{f}(\bx)  \approx  \mathbf{r} \bdot \grad
      \OL{f}(\bx) + \dots
\lb{eq:TaylorExpand}\end{equation}
where we neglect higher order terms.

Substituting the first term in the Taylor expansion of each of $\delta \OL{\rho}$ and $\delta \OL{\bu}$ into eq. \eqref{eq:SubgridTauDeltas} gives
\begin{eqnarray}
  \overline{\tau} (\overline{\rho}, \overline{\bu}_i)
  &=& \left(\partial_k\OL\rho\right)\left(\partial_m\OL{u}_i\right) \left[\left\langle r_k \,r_m  \right\rangle_{\ell} - \left\langle r_k \right\rangle_{\ell} \left\langle r_m  \right\rangle_{\ell}\right]\nonumber\\
  &=& \left(\partial_k\OL\rho\right)\left(\partial_m\OL{u}_i\right) \left[\frac{1}{3}\delta_{km}\,\ell^2\int \mathrm{d}^3 \mathbf{r} \, G (\mathbf{r}) \left | \mathbf{r}\right |^2  \right]\\\nonumber
  &=& \frac{1}{3} \, \ell^2 \, C_2 \,\partial_k \overline{\rho} \, \partial_k \overline{u}_i\end{eqnarray}
This is the nonlinear model of the subscale mass flux $\overline{\tau}_{\ell} (\rho, u_i)$. In deriving the second line, we used  the symmtery of the kernel such that $\left\langle r_k \right\rangle_{\ell}=0$. In the final expression, $C_2 = \int \mathrm{d}^3 \mathbf{r} \, G (\mathbf{r}) \left | \mathbf{r}
\right |^2$  depends solely on the shape of kernel $G$ and, in particular, is independent of scale $\ell$.

We now have a nonlinear model of $\Lambda_\ell$ that is only a function of filtered fields, which are resolved in LES simulations.
We can use this model to gain insight into the mechanism by which $\Lambda$ 
transfers energy across scales. The velocity
gradient tensor $\partial_k \overline{u}_j$ can be decomposed into symmetric
and antisymmetric parts
\begin{equation}
  \partial_i \overline{u}_j = \overline{S}_{ij} + \overline{\Omega}_{ij},
\end{equation}
with
\begin{equation}
\begin{split}
  \overline{S}_{ij} &= \frac{1}{2} \, \left (\partial_i \overline{u}_j
    + \partial_j \overline{u}_i \right ) \\
      \overline{\Omega}_{ij} &= \frac{1}{2} \, \left (\partial_i \overline{u}_j
    - \partial_j \overline{u}_i \right )
    = \frac{1}{2} \, \epsilon_{ijk} \overline{\omega}_k,
\end{split}
\end{equation}
where ${{\bomega}} = \grad \btimes \bu$ is vorticity and $\epsilon_{ijk}$ is the Levi-Civita symbol.
Therefore, $\Lambda$ at any scale $\ell$ can be approximated by a nonlinear model, $\Lambda_{\mModel}$, everywhere in space:
\begin{equation}
\label{eq:lambda_decomp}
\begin{split}
  \Lambda (\mathbf{x}) \approx \Lambda_{\mModel} (\mathbf{x}) &= 
  \frac{1}{3} \, C_2 \, \ell^2 \,
      \frac{1}{\overline{\rho}} \, \left (\partial_j \overline{P} ~
      \partial_k \overline{\rho} ~ \partial_k \overline{u}_j
 \right ) \\
   &= \frac{1}{3} \, C_2 \, \ell^2 \, \frac{1}{\overline{\rho}} \, \left [
      \grad \overline{P} \bdot \overline{\bS}\bdot \grad \overline{\rho} 
    + \frac{1}{2} \, \overline{\bomega} \bdot \left (\grad \overline{\rho}
      \btimes \grad \overline{P} \right ) \right ] \\
   &= \Lambda_{SR} + \Lambda_{BC}
\end{split}
\end{equation}
where 
\be\Lambda_{SR} = \frac{1}{3} \, C_2 \, \ell^2 \, \frac{1}{\overline{\rho}} \, \left [
      \grad \overline{P} \bdot \overline{\bS}\bdot \grad \overline{\rho} 
     \right ]
\lb{eq:Lambda_SR}\ee
is the strain generation process of baropycnal work and
\be\Lambda_{BC} = \frac{1}{3} \, C_2 \, \ell^2 \, \frac{1}{\overline{\rho}} \, \left [
       \frac{1}{2} \, \overline{\bomega} \bdot \left (\grad \overline{\rho}
      \btimes \grad \overline{P} \right ) \right ]
\lb{eq:Lambda_BC}\ee
is its baroclinic vorticity generation process. Equation \eqref{eq:lambda_decomp} is the main result of this paper. In the following sections, we will provide numerical support showing excellent pointwise agreement between $\Lambda$ and its nonlinear model $\Lambda_{\mModel}$.

To illustrate how strain generation by baropycnal work takes place, consider an unstably stratified flow configuration in which $\grad\OL{P}$ and $\grad\OL{\rho}$ in $\Lambda_{SR}$ are anti-aligned ($\grad\OL{\rho}\bdot\grad\OL{P}<0$) as illustrated in Fig. 1 of \cite{Aluie13}, and both are parallel to a contracting eigenvector of $\OL\bS$ (associated with a negative eigenvalue of  $\OL\bS$). Remember that the strain $\OL\bS$ in our nonlinear model arises from $\OL\tau_\ell(\rho,\bu)$ which represents scales smaller than $\ell$.
In such a configuration, the contraction (and therefore strain) is enhanced leading to the generation of kinetic energy in the form of straining motion at scales smaller than $\ell$ ($\Lambda_\ell>0$ in eq. \eqref{largeKE}). The ultimate source of kinetic energy being transferred by $\Lambda$ to motions at scales $<\ell$ is potential energy due to the large-scale pressure gradient, $\grad\OL{P}$.

The baroclinic component, $\Lambda_{BC}$, demonstrates how baroclinicity, $\grad \rho \btimes \grad P$, 
plays a role in the energetics across scales.
The importance of baroclinicity is well known \cite{Sharp84,KunduCohen08} but it has always been analyzed within the vorticity budget. Its contribution to the energy budget has never been clear.
Baroclinicity in the kinetic energy budget arises naturally from our scale decomposition and the identification of $\Lambda$ as a scale-transfer mechanism. The need for a scale decomposition in order for $\Lambda$ and, as a result, baroclinic energy transfer, to appear in the kinetic energy budget should not be surprising. This is similar to the scale transfer term $\Pi$, which does not appear in the budget without disentangling scales due to energy conservation. In the same vein, the appearance of baroclinicity in the vorticity equation can be interpreted as being a consequence of an effective scale decomposition performed by 
the curl operator $\grad\btimes$, which a high-pass filter.

As mentioned in the introduction, in the compressible LES literature,
$\Lambda$ is almost always lumped with pressure-dilatation, $\OL{P}\grad\bdot\OL{\bu}$ in the form of $\OL{P}\grad\bdot\wt{\bu}$ \cite{Favre69,Lele94,Garnieretal09,OBrienetal2014}, thereby completely missing the physical processes inherent in baropycnal work. Our analysis here supports the argument in \cite{Aluie11,Aluie13} to separate $\Lambda$ from pressure-dilatation. In those studies, it was reasoned that $\Lambda$ and $\OL{P}\grad\bdot\OL{\bu}$ are fundamentally different; the former involves interactions between the large scale pressure gradient with subscale fluctuations, allowing the transfer energy across scales, whereas the latter is solely due to large-scale fields and cannot participate in the transfer of energy \emph{across} scales.

\section{Simulations}\lb{sec:simulations}
To provide empirical  support to our nonlinear model of $\Lambda$, we carry out a suite of DNS 
of forced compressible turbulence in a periodic box of size $2 \pi$ on which we perform \textit{a priori} tests of our derived model against simulation data. The DNS solve the fully compressible Navier Stokes equations:
\begin{eqnarray} 
&\hspace{-0.4cm}\partial_t \rho& + \partial_j(\rho u_j) = 0 \lb{continuity} \\
&\hspace{-0.4cm}\partial_t (\rho u_i)& + \partial_j(\rho u_i u_j) 
= -\partial_i P +  \partial_j\sigma_{ij} + \rho F_i   \lb{momentum}\\
&\hspace{-0.4cm}\partial_t (\rho E)&+ \partial_j(\rho E u_j) 
= -\partial_j (P u_j) +\partial_j[2\mu ~ u_i(S_{ij} - \frac{1}{d} S_{kk}\delta_{ij})]  -\partial_j q_j  +\rho u_i F_i - \mathcal{RL} \lb{total-energy}\end{eqnarray}
Here, $\bu$ is velocity, $\rho$ is density, $E=|\bu|^2/2 + e$ is total energy per unit mass, where $e$ is specific internal energy,  $P$ is thermodynamic pressure, $\mu$ is dynamic viscosity, ${\bf q} = -\kappa \grad T$ is the heat flux with a thermal conductivity $\kappa$ and temperature $T$.  Both dynamic viscosity and thermal conductivity are spatially variable, where $\mu(\bx) = \mu_0 (T(\bx)/T_0)^{0.76}$. Thermal conductivity is set to satisfy a Prandtl number $Pr = c_p \mu/\kappa =0.7$, where $c_p=R\,\gamma/(\gamma-1)$ is the specific heat with specific gas constant $R$ and $\gamma=5/3$. We use the ideal gas equation of state, $P=\rho R T$. 
We stir the flow using an external acceleration field $F_i$, and $\mathcal{RL}$ represents radiation losses from internal energy. 
$S_{ij} = (\partial_j u_i + \partial_i u_j)/2$ is the symmetric strain tensor and $\sigma_{ij}$ is the the deviatoric (traceless) viscous stress 
\be \sigma_{ij}=2\mu(S_{ij} - \frac{1}{3} S_{kk}\delta_{ij})
\lb{eq:viscousstress}\ee

We solve the above equations using the pseudo-spectral method with $2/3$rd dealiasing. We advance in time using the $4^{th}$-order Runge-Kutta scheme with a variable time step.

The acceleration $\bF$ we use is similar to that in \cite{Federrathetal08}. 
In Fourier space, the acceleration is defined as
\begin{equation}
  \widehat{F}_i(\mathbf{k}) = \widehat{f}_j (\mathbf{k})
    P_{ij}^\zeta (\mathbf{k}),
\end{equation}
where the complex vector $\mathbf{\widehat{f}}$ is constructed from 
independent Ornstein-Uhlenbeck stochastic processes \cite{EswaranPope88} and the
projection operator $P_{ij}^\zeta(\mathbf{k}) = \zeta \delta_{ij} +
(1 - 2\zeta)\frac{k_i k_j}{|\mathbf{k}|^2}$ allows to control the ratio of
solenoidal ($\nabla \cdot \mathbf{F} = 0$) and dilatational
($\nabla \times \mathbf{F} = 0$) components of the forcing using the parameter
$\zeta$. When $\zeta = 0$, the forcing is purely dilatational and when $\zeta = 1$,
the forcing is purely solenoidal. The acceleration is constrained to low wavenumbers 
$|\mathbf{k}| < k_F$.

\begin{figure}[h]
\center{\includegraphics[width=0.5\textwidth]
    {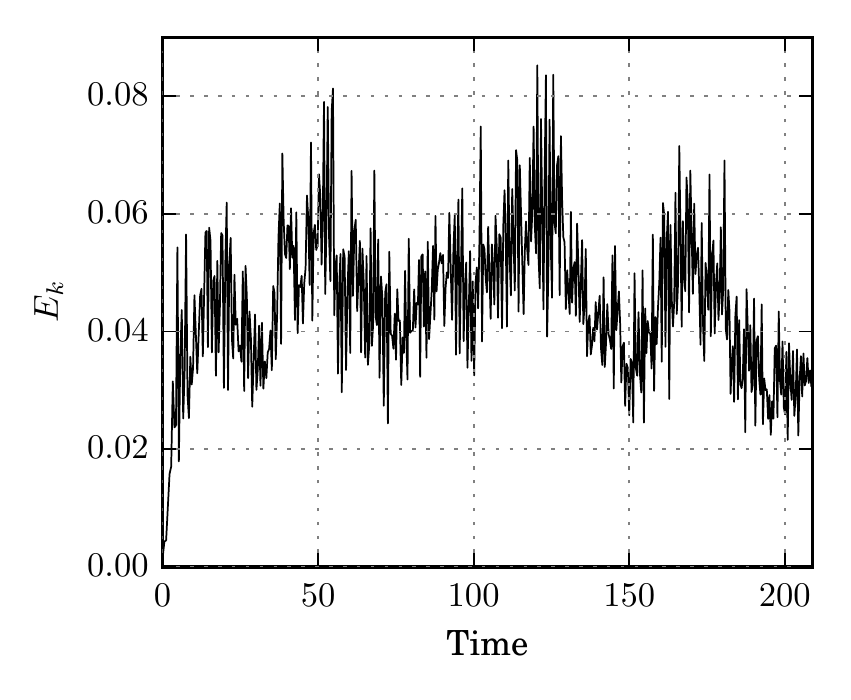}}
\caption{\label{fig:Ek_timeseries} Time-series of average kinetic energy showing a statistically steady state.}
\end{figure}

For the internal energy loss term $\mathcal{RL}$, we have tested two schemes: a spatially varying radiative loss, $\mathcal{RL}=\rho \bu\bdot\bF$, and another that is independent of space, $\mathcal{RL}=\langle\rho \bu\bdot\bF\rangle$. The two schemes yield indistinguishable results. Including an internal energy loss term is similar to what was done in other studies \cite{JagannathanDonzis16,Wangetal10} to allow total energy to remain stationary. After an initial transient, mean kinetic energy reaches a statistically stationary state as shown in Fig. \ref{fig:Ek_timeseries}.

\begin{table*}
\caption{\label{tab:sim_params} Comparison of compressibility metrics and
  cascade terms at different grid resolution. Low-$\zeta$ corresponds to high compressibility in the external forcing. 
  The spatially averaged cascade terms $\langle\Lambda_{\ell}\rangle$
  and $\langle\Pi_{\ell}\rangle$ are calculated using the sharp-spectral cutoff filter with
  $k_{\ell} = 2\pi/\ell=6$.
  $\frac{\Delta x}{\eta}$ is the ratio of grid size to the Kolmogorov length.}
\centering
\begin{tabular}{cccccccccc}
\toprule
  \textrm{Run} &
  \multicolumn{1}{c}{$N$} &
  \multicolumn{1}{c}{$\zeta$} &
  \multicolumn{1}{c}{$M_t$} &
  \multicolumn{1}{c}{$Re_{\lambda}$} &
  \multicolumn{1}{c}{$\frac{(\grad \bdot \bu)_{\scriptsize\mbox{rms}}}
    {(\grad \btimes \bu)_{\scriptsize\mbox{rms}}}$} &
  \multicolumn{1}{c}{$\frac{K^d}{K^s}$} &
  \multicolumn{1}{c}{$\frac{\Delta x}{\eta}$} &
  \multicolumn{1}{c}{$\left <\Pi_{\ell} \right \rangle$} &
  \multicolumn{1}{c}{$\left <\Lambda_{\ell} \right \rangle$} \\
\toprule
    1 & 1024 & 0.01 & 0.23 & 65  & 0.50  & 0.74 & 0.23 &
        $8.9\times10^{-3}$ & $-5.6\times10^{-3}$ \rule{0pt}{2.9ex} \\
    2 & 512  & 0.01 & 0.22 & 33  & 0.46  & 0.51 & 0.28 &
        $6.8\times10^{-3}$ & $-4.8\times10^{-3}$ \\
    3 & 512  & 0.6  & 0.33 & 206 & 0.05  & 0.02 & 1.78 &
        $2.3\times10^{-2}$ & $-9.6\times10^{-5}$ \\
    4 & 256  & 0.01 & 0.21 & 18  & 0.54  & 0.56 & 0.25 &
        $3.6\times10^{-3}$ & $-3.1\times10^{-3}$ \\
    5 & 256  & 0.6  & 0.42 & 150 & 0.04  & 0.01  & 2.1 &
        $5.0\times10^{-2}$ & $-6.0\times10^{-4}$ \\
    6 & 256  & 1.0  & 0.46 & 175 & 0.03  & 0.003 & 2.2 &
        $5.0\times10^{-2}$ & $-3.8\times10^{-4}$ \\
    7 & 128  & 0.01 & 0.20 & 10  & 0.65  & 0.80 & 0.24 &
        $8.0\times10^{-4}$ & $-7.5\times10^{-4}$ \\
    8 & 128  & 0.6  & 0.50 & 105 & 0.05  & 0.01 & 2.3  &
        $4.0\times10^{-2}$ & $-4.0\times10^{-4}$ \\
    9 & 128  & 1.0  & 0.40 & 95  & 0.03  & 0.01 & 2.0  &
        $2.5\times10^{-2}$ & $-2.0\times10^{-4}$ \\
\toprule
\end{tabular}
\end{table*}
Table \ref{tab:sim_params} summarizes the simulations we ran for this study and various metrics
characterizing the importance of compressibility effects in each run. The
turbulent Mach number is $M_t = \left \langle u_i u_i \right \rangle^{1/2} /\left \langle c \right\rangle$
and the Taylor Reynolds number is $Re_{\lambda} = \left \langle(u_i u_i)/3 \right \rangle^{1/2} \lambda/
 \left \langle \mu/\rho \right \rangle$. Here $c = \sqrt{\gamma p/\rho}$
is the sound speed and $\lambda = \left \langle u_i u_i \right \rangle^{1/2} /
\left  \langle u_{i,i}^2 \right \rangle^{1/2}$ is the Taylor microscale. The compressibility
metrics in Table \ref{tab:sim_params} show the relative importance of dilatational versus solenoidal velocity
modes. We use the Helmholtz decomposition, $\bu=\bu^d + \bu^s$ to
obtain the dilatational ($\nabla \times \mathbf{u}^d = 0$) and solenoidal
($\nabla \cdot \mathbf{u}^s = 0$) components of the velocity field. The
dilatational kinetic energy is $K^d = \left \langle \rho u_i^d u_i^d / 2 \right \rangle$
and the solenoidal kinetic energy is $K^s = \left \langle \rho u_i^s u_i^s / 2
\right \rangle$. Their ratio $K^d/K^s$ yields a measure of compressibility at 
large scales.  It is well-known \cite{Federrathetal08,PetersenLivescu10} that the 
dilatational kinetic energy $K^d$ becomes significant when forced 
directly using $\bF$ with a small $\zeta$. This holds, even though 
the low-$\zeta$ runs have a lower Mach number than high-$\zeta$ runs.
The ratio 
$(\grad \bdot \bu)_{\scriptsize\mbox{rms}} /(\grad \btimes \bu)_{\scriptsize\mbox{rms}}$
yields a measure of compressibility at small scales. 

The last two columns in table \ref{tab:sim_params} summarize the effect
of $\zeta$ on the relative
importance of $\Pi$ and $\Lambda$. While the deformation work $\Pi$ is
significant in all cases, baropycnal work $\Lambda$, which arises only in variable density flows, 
is greatly affected by the type of forcing used. At the Reynolds numbers we simulate, we find that
$\Lambda$ becomes important only for low $\zeta$ when the dilatational modes are directly forced.
Even at relatively high Mach numbers, $\Lambda$ remains small for high $\zeta$. We caution, however, 
that these observations might be Reynolds number dependent. Moreover, $\Lambda$ has been shown to dominate in non-dilatational low Mach number variable density flows \cite{Livescuetal09,Zhaoetal19}.

\begin{figure}[h]
\begin{subfigure}[b]{\sfigsize}
  \includegraphics[width=\sfigsize]
    {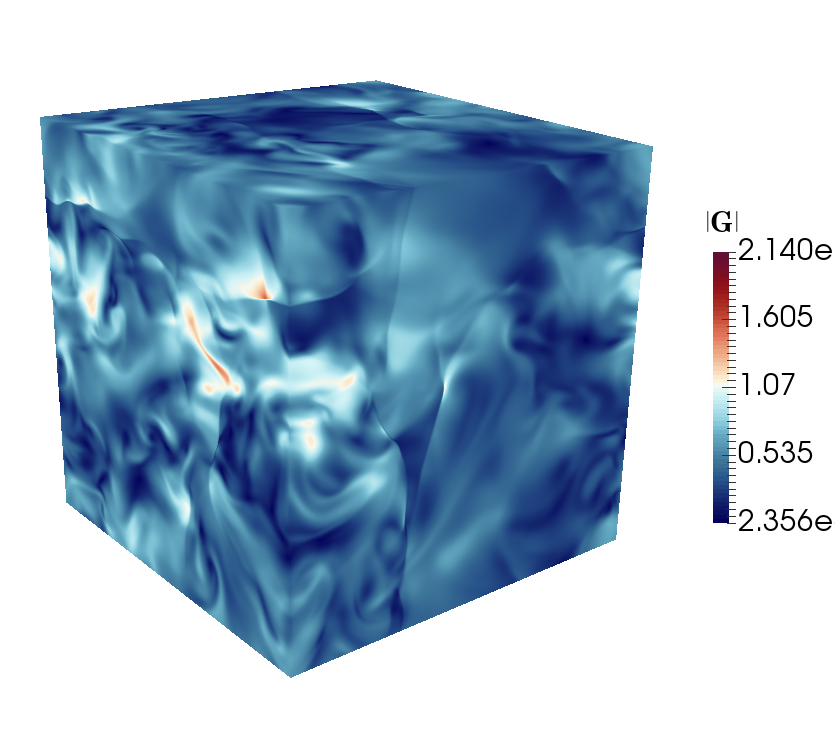}
    \caption{}
    \label{fig:z001_3dview}
\end{subfigure}
\begin{subfigure}[b]{\sfigsize}
  \includegraphics[width=\sfigsize]
    {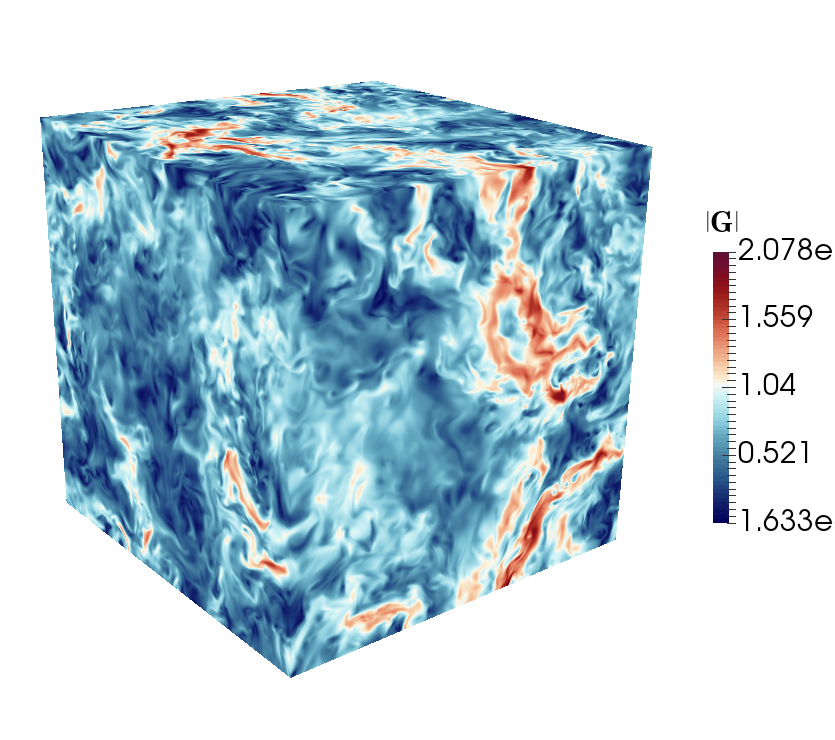}
    \caption{}
    \label{fig:z06_3dview}
\end{subfigure}
\caption{\label{fig:Gm_3dview} Momentum magnitude $|\rho \mathbf{u}|$ from
    (\subref{fig:z001_3dview})  Run 1 ($\zeta = 0.01$) 
    and (\subref{fig:z06_3dview})  Run 3 ($\zeta = 0.6$).}
\end{figure}

Figure \ref{fig:Gm_3dview} shows typical visualizations of the flows 
arising from low-$\zeta$ and high-$\zeta$ forcing. The stark qualitative difference
shows the significance of dilatational forcing on the flow \cite{Federrathetal08,Kritsuketal10}, 
at least in limited resolution simulations. It has been argued \cite{Sarkaretal91,Shivamoggi97,JagannathanDonzis16} that at sufficiently high Reynolds numbers,
flows forced dilatationally will produce sufficient vortical motion to resemble those forced solenoidally.
Since we are primarily interested in the $\Lambda$ term here, unless stated otherwise, plots that follow will be from
low-$\zeta$ simulations where $\Lambda$ is significant.

\begin{figure}[h]
\center{\includegraphics[width=\sfigsize]
    {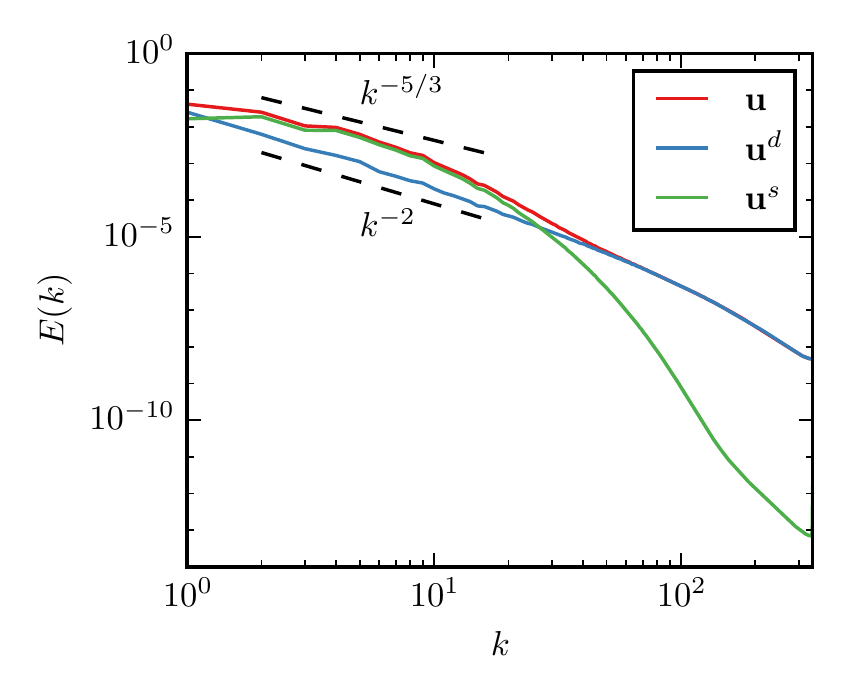}}
\caption{\label{fig:Ek_spec} Spectra of velocity ($u$) and its
    dilatational and solenoidal ($u^d$ and $u^s$, respectively) components from Run 1.
    The reference dashed black lines have slopes of $-5/3$ and $-2$.}
\end{figure}

Figure \ref{fig:Ek_spec} shows the velocity spectra in the case of
highly compressive (low-$\zeta$) forcing. At intermediate scales, the
spectrum of $u^d$ seems to follow a power law close to $k^{-2}$, which is
expected for the dilatational component \cite{PetersenLivescu10,Federrathetal10,Wangetal13}.
It is well-known \cite[e.g.][]{Federrathetal10,Wangetal13} that obtaining a clear power-law scaling of the 
solenoidal velocity is challenging when forcing dilatationally, even at our $1{,}024^3$ resolution.

\begin{figure}[h]
\center{\includegraphics[width=\sfigsize]
    {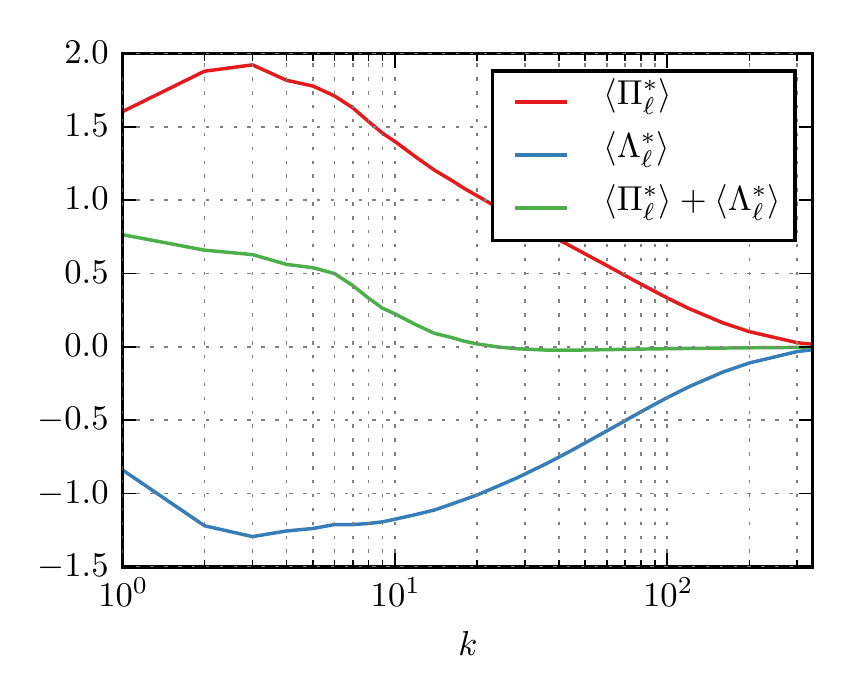}}
\caption{\label{fig:cascade_terms} Flux terms $\Pi_{\ell}$ and
  $\Lambda_{\ell}$ from Run 1, as well as their sum averaged over space and time, as a function of the filtering
  wavenumber $k=2\pi/\ell$. Filtering here uses the sharp-spectral filter kernel.
  The star superscript indicates normalization by the effective kinetic energy injection, $\varepsilon^{eff} =
  \varepsilon^{inj} + \langle p \grad\bdot\bu\rangle$.}
\end{figure}

Figure~\ref{fig:cascade_terms} shows the cascade terms $\Pi_{\ell}$ and
$\Lambda_{\ell}$ averaged over the domain as a function of the
filter wavenumber $k$.  As~is the case in 3D isotropic incompressible
turbulence, $\Pi_{\ell}$ is positive for all wavenumbers, transferring kinetic
energy from large to small scales. On~the other hand, $\Lambda_{\ell}$ 
is negative, effectively reducing the total amount of energy transferred 
across scales. This is consistent with previous studies which 
measured $\Lambda_{\ell}$ in homogeneous isotropic 
compressible turbulence~\cite{Wangetal13}. 
Across a shock, the~pressure and density gradients have the same direction and 
are aligned with the contracting strain eigenvector, 
leading to negative baropycnal work~\cite{Aluie13,EyinkDrivas18}, thereby 
reducing the  intensity of the cascade. 
Using the terminology of \cite{AlexakisBiferale18},
this is a ``bi-directional cascade.''
The~situation is different in buoyancy 
driven (unstably stratified) flows, where pressure and density gradients are 
in opposite directions leading to positive baropycnal work~\cite{Aluie13}. 
Within the framework of Reynolds-averaged Navier-Stokes (RANS), it has been shown that for
variable density flows, such as turbulence generated by the Rayleigh-Taylor
instability, the~RANS equivalent of $\Lambda$ is the largest contributor to the
kinetic energy cascade~\cite{Livescuetal09}. We shall
present our results on $\Lambda$ in buoyancy driven flows in forthcoming work~\cite{Zhaoetal19}.

\clearpage
\newpage
\section{Numerical Results}
\begin{figure}[h]
\begin{subfigure}[b]{\sfigsize}
  \includegraphics[width=\sfigsize]
    {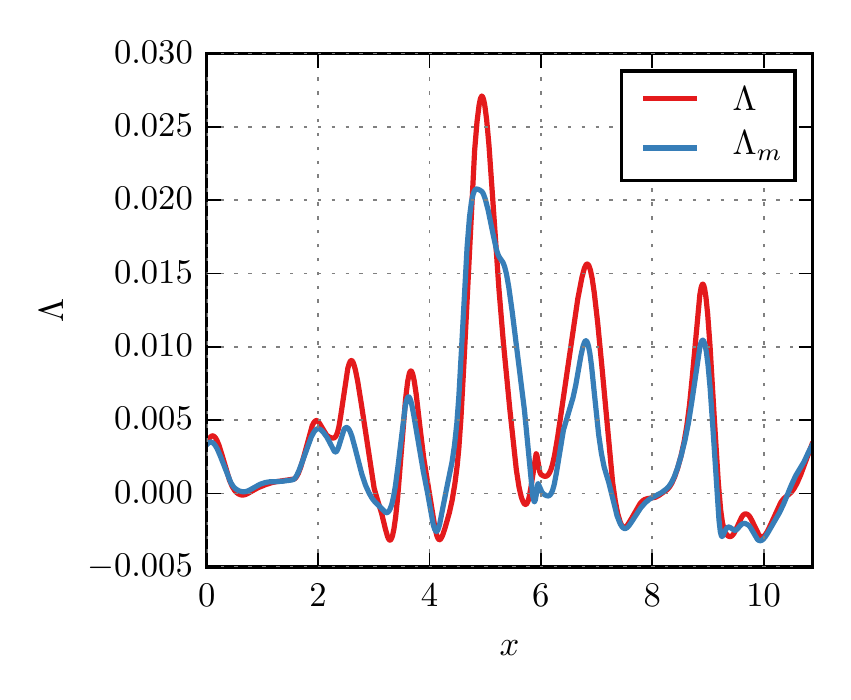}
    \caption{}
    \label{fig:plotoverline_nl_boxk8}
\end{subfigure}
\begin{subfigure}[b]{\sfigsize}
  \includegraphics[width=\sfigsize]
    {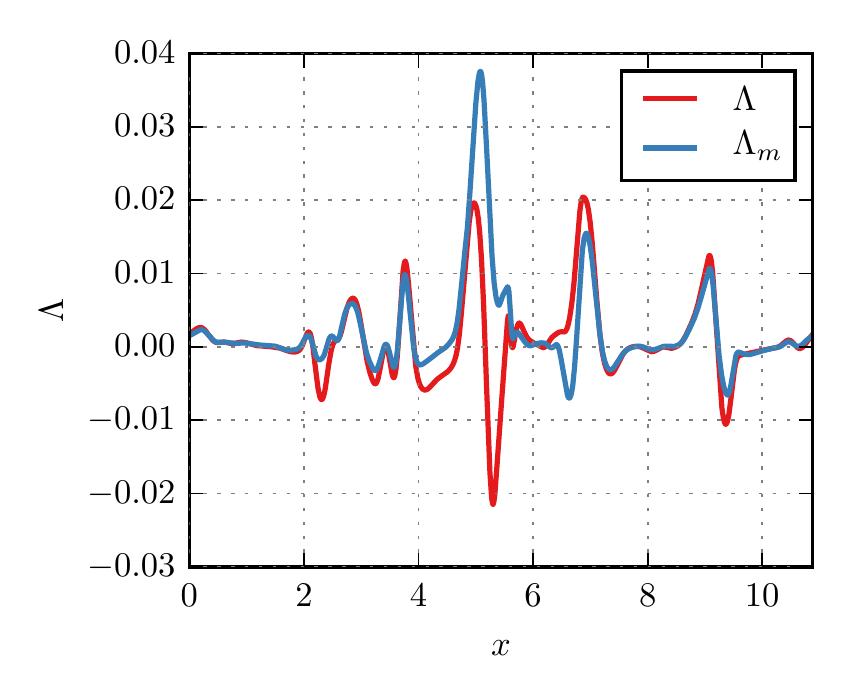}
    \caption{}
    \label{fig:plotoverline_nl_boxk16}
\end{subfigure}
\begin{subfigure}[b]{\sfigsize}
  \includegraphics[width=\sfigsize]
    {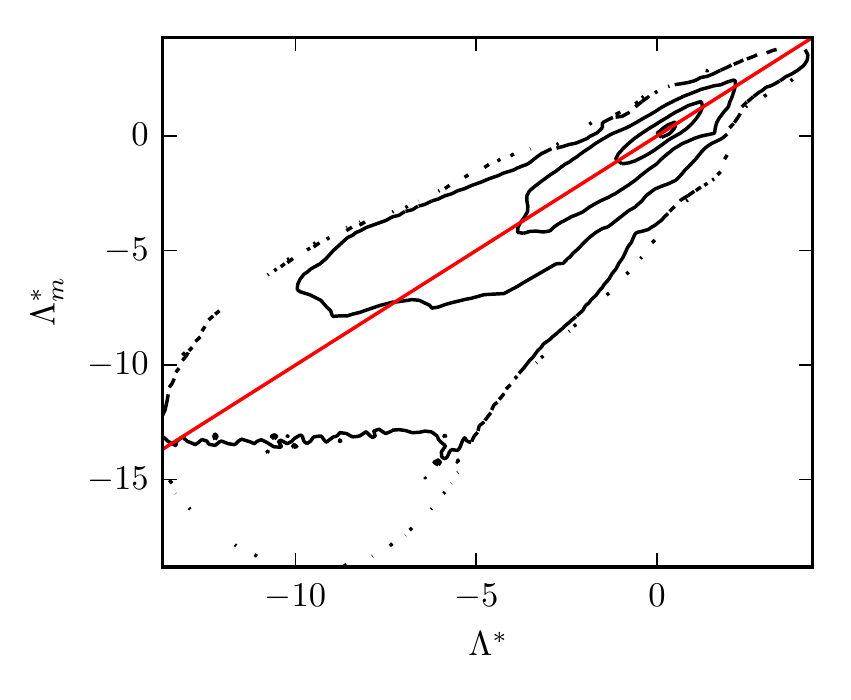}
    \caption{}
    \label{fig:jpdf_nl_boxk8}
\end{subfigure}
\begin{subfigure}[b]{\sfigsize}
  \includegraphics[width=\sfigsize]
    {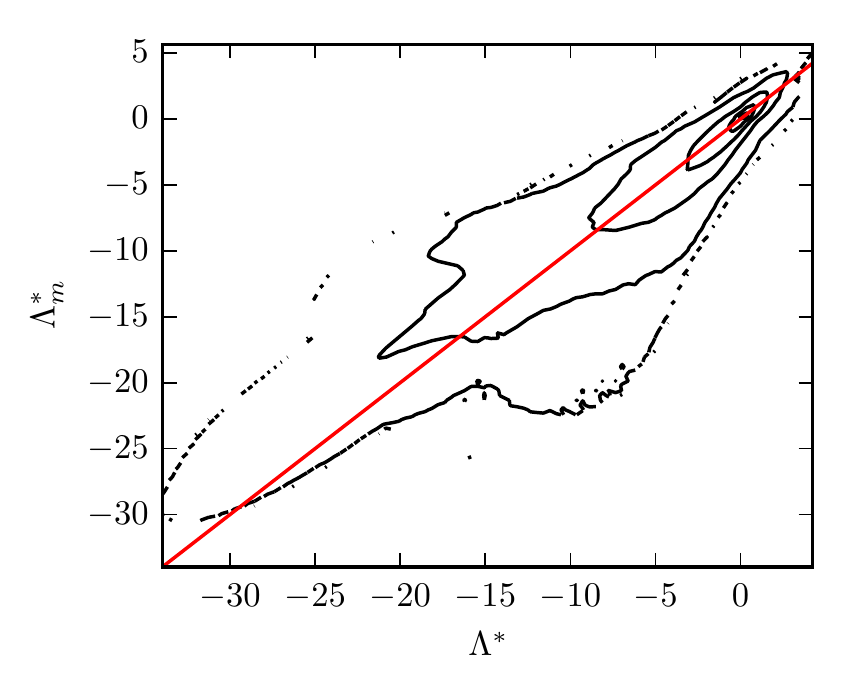}
    \caption{}
    \label{fig:jpdf_nl_boxk16}
\end{subfigure}
\caption{\label{fig:nl_model_box} Correlation between baropycnal work and its nonlinear model
from Run 1 using the box kernel with filter scale 
$k_{\ell} = 8$ in (\subref{fig:plotoverline_nl_boxk8},\subref{fig:jpdf_nl_boxk8}) and 
$k_{\ell} = 16$ in (\subref{fig:plotoverline_nl_boxk16},\subref{fig:jpdf_nl_boxk16}).
Top two panels plot $\Lambda$ and $\Lambda_{\mModel}$ along a diagonal line through the 
domain from a single snapshot.
Lower two panels show time-averaged isocontours of the logarithm of the joint PDF between $\Lambda$ and $\Lambda_{\mModel}$,
where star superscripts indicate that means have been subtracted and the values are normalized by their variance. 
Straight-red lines are $y=x$.
The correlation coefficients are $R_c = 0.93$ at filter scale $k_{\ell} = 8$ and $R_c = 0.94$ at $k_{\ell} = 16$. All four panels indicate excellent correlation between $\Lambda$ and $\Lambda_{\mModel}$.}
\end{figure}

\begin{figure}[h]
\begin{subfigure}[b]{\sfigsize}
  \includegraphics[width=\sfigsize]
    {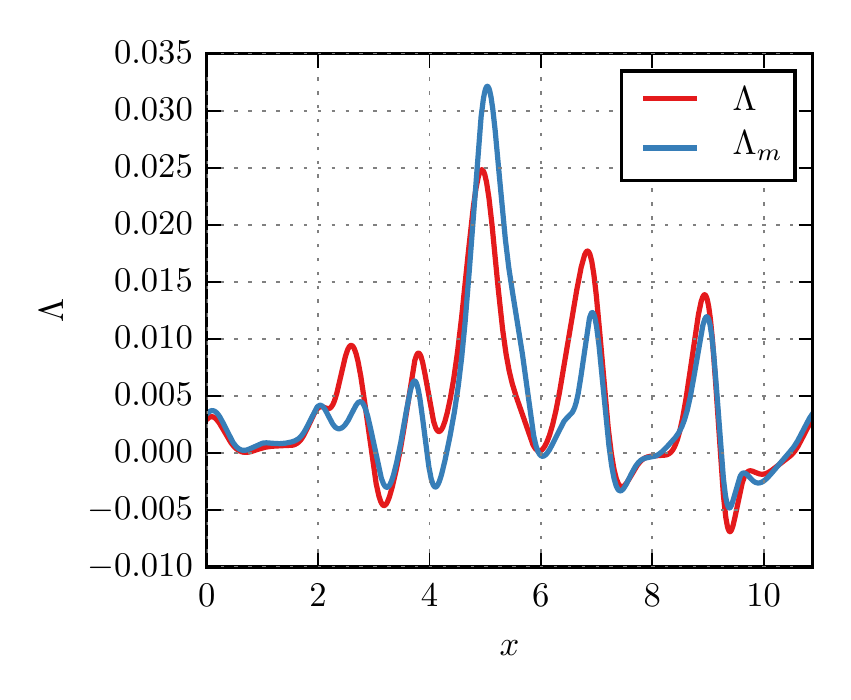}
    \caption{}
    \label{fig:plotoverline_nl_gaussk8}
\end{subfigure}
\begin{subfigure}[b]{\sfigsize}
  \includegraphics[width=\sfigsize]
    {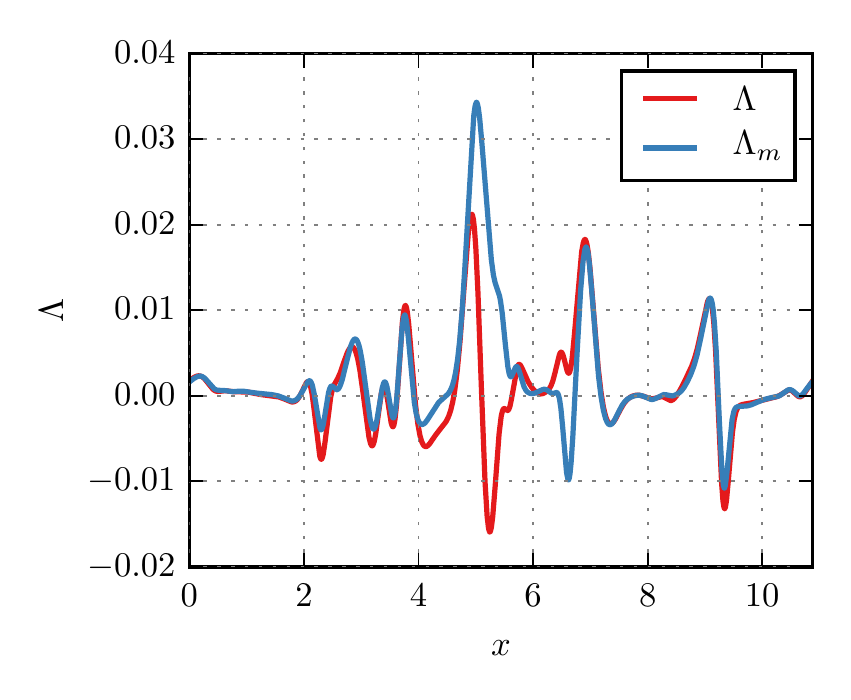}
    \caption{}
    \label{fig:plotoverline_nl_gaussk16}
\end{subfigure}
\begin{subfigure}[b]{\sfigsize}
  \includegraphics[width=\sfigsize]
    {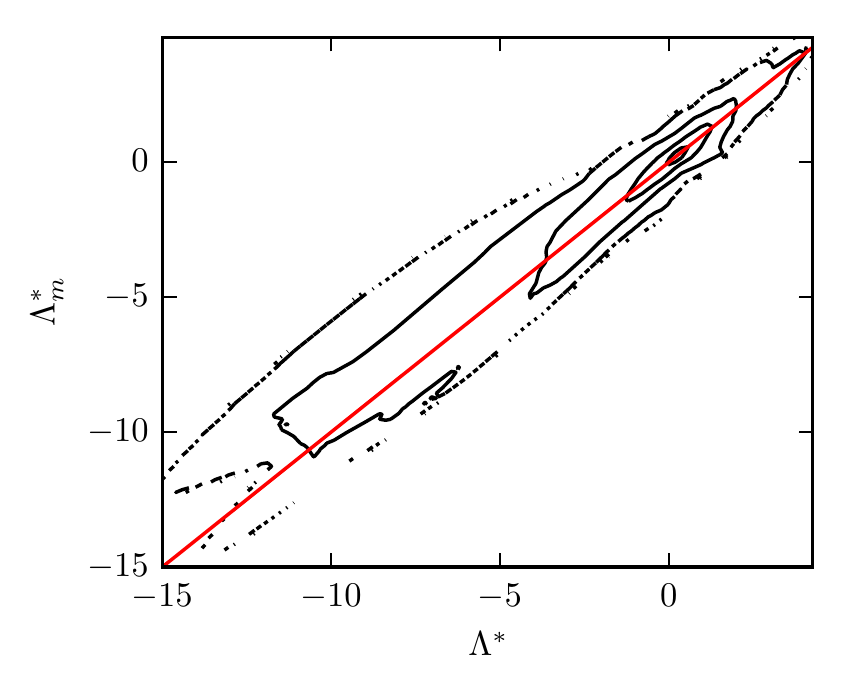}
    \caption{}
    \label{fig:jpdf_nl_gaussk8}
\end{subfigure}
\begin{subfigure}[b]{\sfigsize}
  \includegraphics[width=\sfigsize]
    {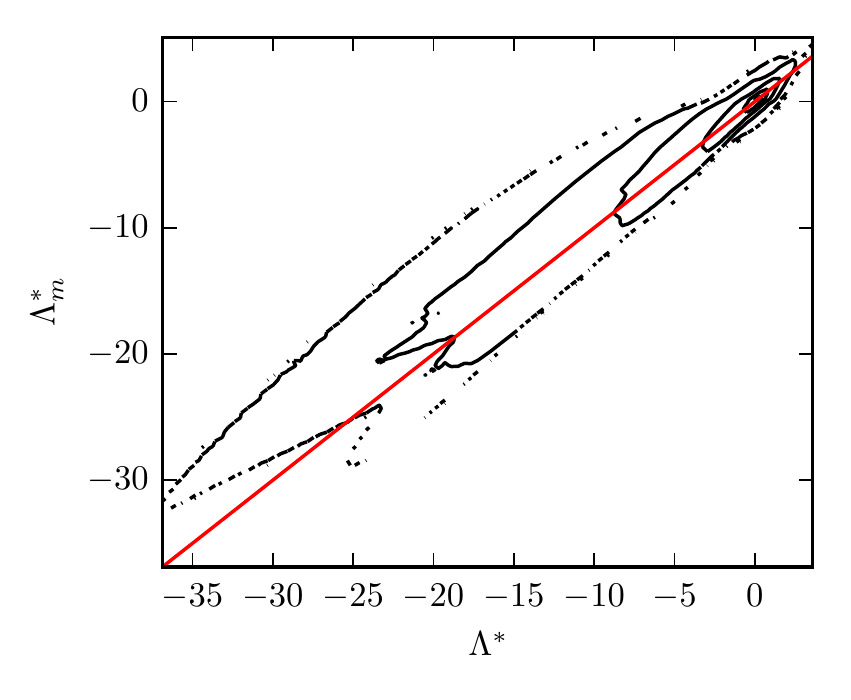}
    \caption{}
    \label{fig:jpdf_nl_gaussk16}
\end{subfigure}
\caption{\label{fig:nl_model_gauss} Same as in Fig. \ref{fig:nl_model_box} but using the Gaussian filter. 
The correlation coefficients are $R_c = 0.97$ at filter scale $k_{\ell} = 8$ and $R_c = 0.97$ at $k_{\ell} = 16$.
All four panels indicate excellent correlation between $\Lambda$ and $\Lambda_{\mModel}$.}
\end{figure}

\begin{figure}[h]
\begin{subfigure}[b]{\sfigsize}
  \includegraphics[width=\sfigsize]
    {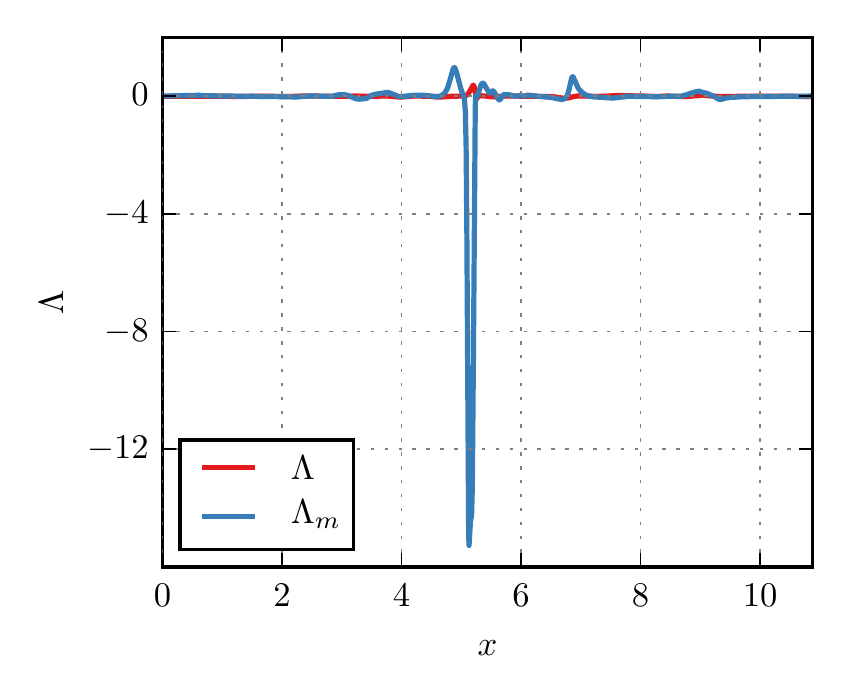}
    \caption{}
    \label{fig:plotoverline_nl_cutoffk8}
\end{subfigure}
\begin{subfigure}[b]{\sfigsize}
  \includegraphics[width=\sfigsize]
    {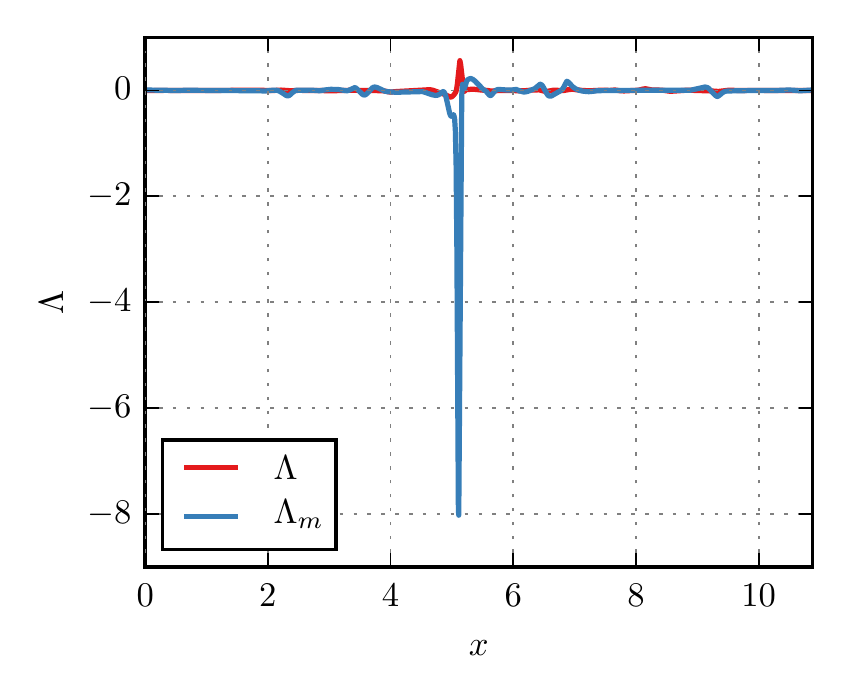}
    \caption{}
    \label{fig:plotoverline_nl_cutoffk16}
\end{subfigure}
\begin{subfigure}[b]{\sfigsize}
  \includegraphics[width=\sfigsize]
    {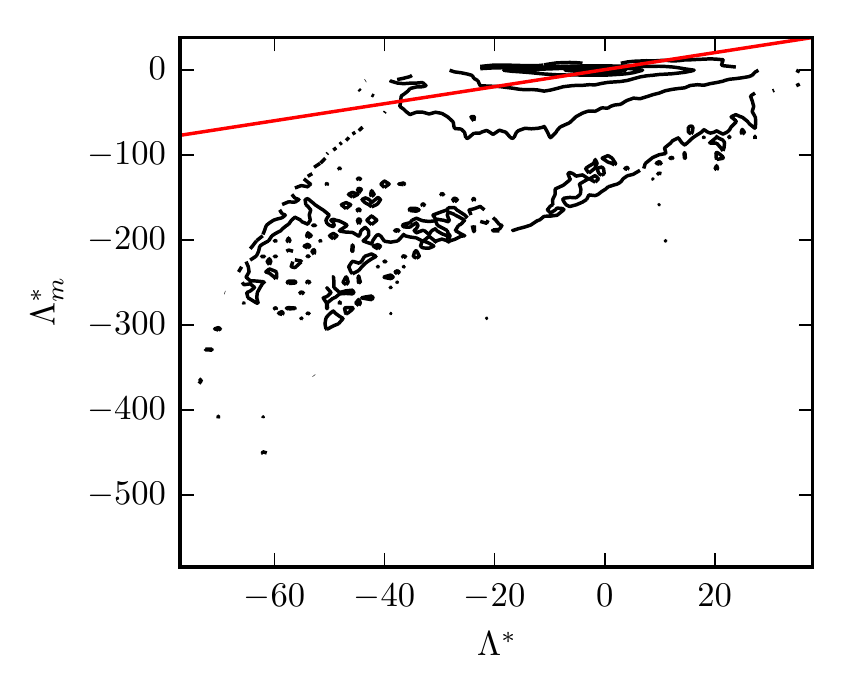}
    \caption{}
    \label{fig:jpdf_nl_cutoffk8}
\end{subfigure}
\begin{subfigure}[b]{\sfigsize}
  \includegraphics[width=\sfigsize]
    {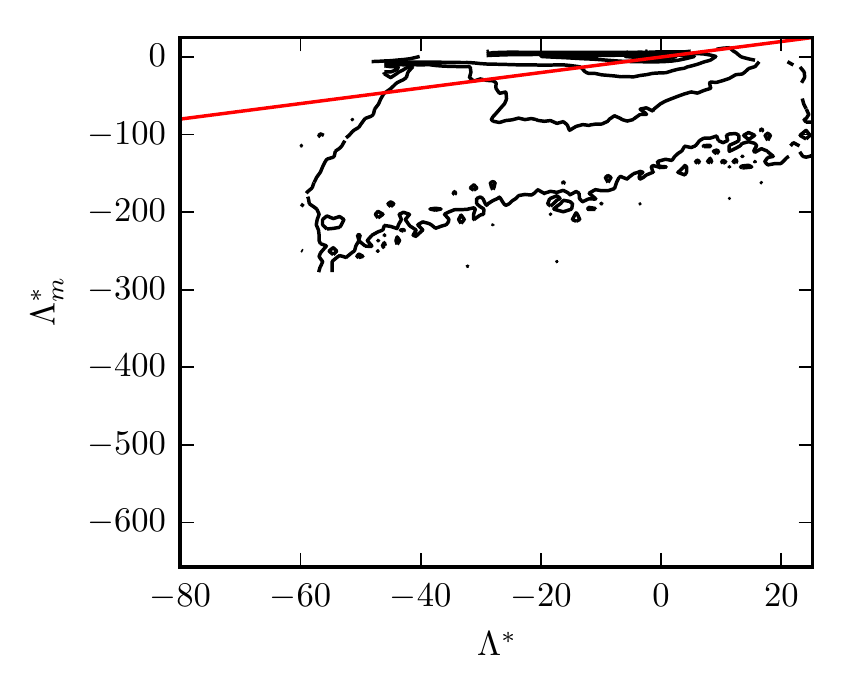}
    \caption{}
    \label{fig:jpdf_nl_cutoffk16}
\end{subfigure}
\caption{\label{fig:nl_model_cutoff} Same as in Fig. \ref{fig:nl_model_box} but using the sharp-spectral filter.
The correlation coefficients are $R_c = 0.27$ at filter scale $k_{\ell} = 8$ and $R_c = 0.28$ at $k_{\ell} = 16$.
The correlation between $\Lambda$ and $\Lambda_{\mModel}$ is poor when using a sharp-spectral filter due to its nonpositivity,
which can yield negative filtered densities \cite{Aluie13} and physically unrealizable stresses \cite{Vremanetal94}.}
\end{figure}

\begin{figure}[h]
\begin{subfigure}[b]{\sfigsize}
  \includegraphics[width=\sfigsize]
    {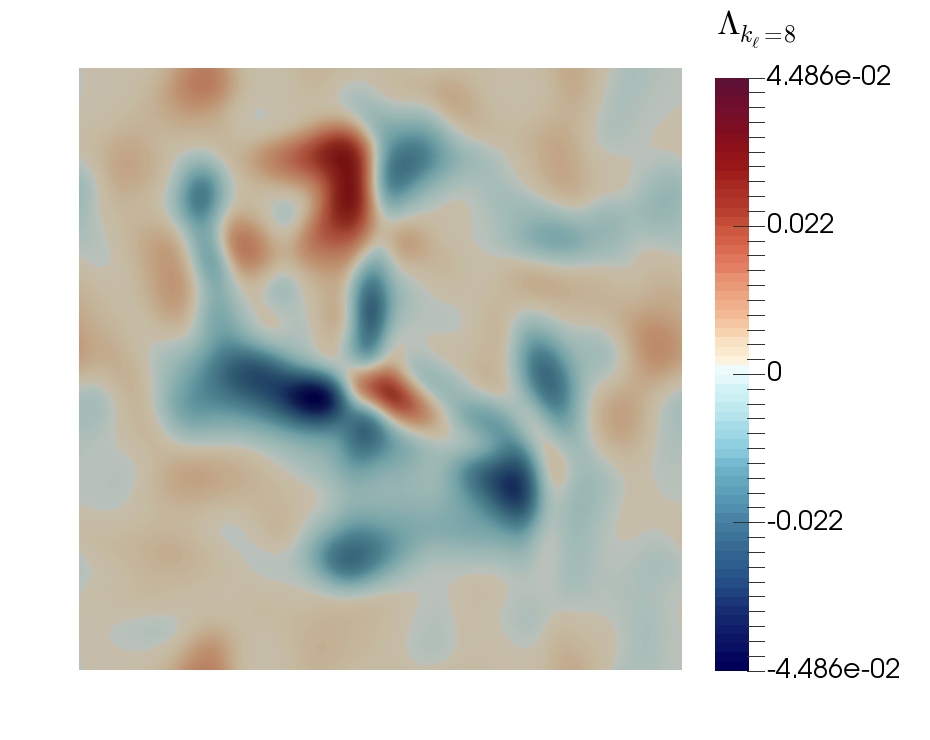}
    \caption{}
    \label{fig:slice_lambda_exact}
\end{subfigure}
\begin{subfigure}[b]{\sfigsize}
  \includegraphics[width=\sfigsize]
    {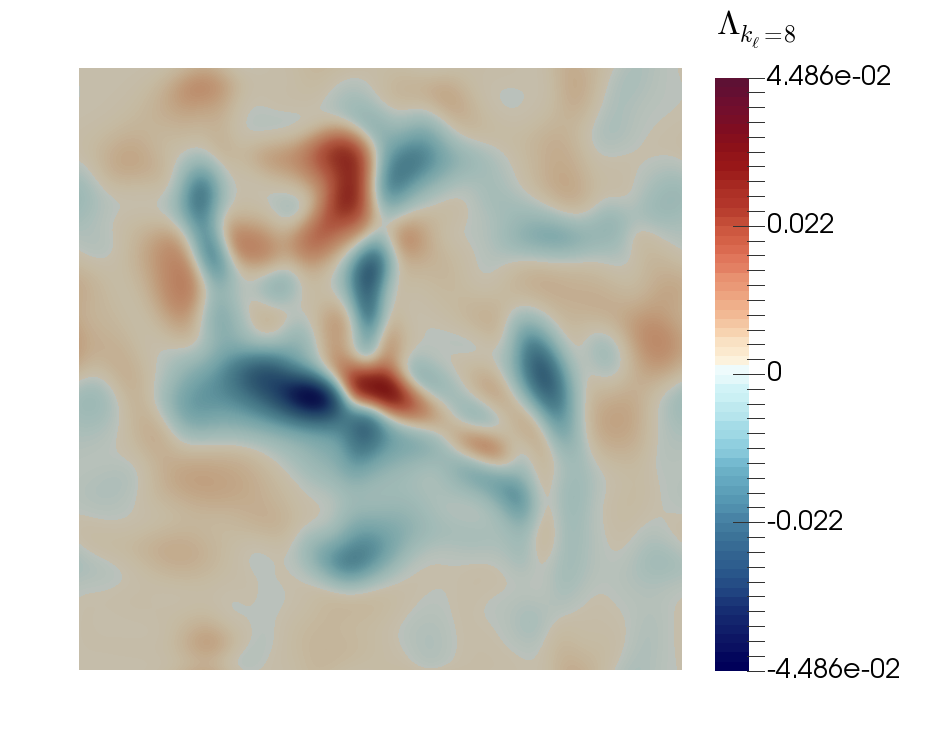}
    \caption{}
    \label{fig:slice_lambda_model}
\end{subfigure}
\caption{\label{fig:nl_model_slice} 
Pointwise comparison between (a) baropycnal work and (b) its nonlinear model 
from a 2D slice of the 3D domain in Run 1, at one instant in time.
A Gaussian kernel at scale $k_{\ell} = 8$ is used. The visualizations show excellent pointwise correlation.}
\end{figure}

\lb{sec:NumericalResults}
To quantify the pointwise agreement between baropycnal work $\Lambda$ and its nonlinear model $\Lambda_{\mModel}$ in our DNS, 
we measure the correlation coefficient:
\begin{equation}
  R_c = \frac{\left \langle \Lambda_{\mModel} \Lambda \right \rangle - \left \langle \Lambda_{\mModel} \right \rangle
      \left \langle \Lambda \right \rangle}
      {\left [ \left ( \left \langle \Lambda_{\mModel}^2 \right \rangle
    - \left \langle \Lambda_{\mModel} \right \rangle^2 \right )
      \left ( \left \langle \Lambda^2 \right \rangle - \left \langle  \Lambda \right \rangle^2 \right )
      \right ]^{1/2}}.
\end{equation}
We also analyze the joint probability density function (PDF) in Figs. \ref{fig:nl_model_box}-\ref{fig:nl_model_cutoff}, and 
visualize $\Lambda$ and $\Lambda_{\mModel}$ in $x$-space in Fig. \ref{fig:nl_model_slice}.

\begin{table}
\setlength{\tabcolsep}{24pt}
\caption{\label{tab:fltrs} The types of filters used in calculating $\Lambda$ and $\Lambda_{\mModel}$. The Heaviside function $H(x)=1$ for $x\ge0$ and $H(x)=0$ for $x<0$. The correlation coefficient $R_c$ is shown at two scales $k_\ell = 2\pi/\ell$.}
\centering
\begin{tabular}{llcc}
\toprule
  \textrm{Filter type} &
  \textrm{Kernel} &
$R_c|\, k_\ell=8$ &
$R_c|\, k_\ell=16$ \\
\toprule
  Box &
    $G_{\ell}(\mathbf{x}) = \prod_{i=1}^3 \frac{1}{\ell}
        H \left (\frac{\ell}{2} - \left | x_i \right | \right )$
        \rule{0pt}{2.9ex} & 0.93 & 0.94 \\
  Gaussian &
    $G_{\ell}(\mathbf{x}) = \frac{1}{\ell^3} \left (\frac{6}{\pi}
        \right )^{3/2} e^{-\frac{6 \left |\mathbf{x} \right |^2}{\ell^2}}$
        \rule{0pt}{4.9ex} & 0.97 & 0.97 \\
  Sharp spectral &
    $\wh{G}_{\ell}(\mathbf{k}) = \prod_{i=1}^3 H \left (\frac{2\pi}{\ell} - \left | k_i \right | \right )$
        \rule{0pt}{4.9ex} & 0.27 & 0.28\\
\toprule
\end{tabular}
\end{table}

In our study, we use the filters defined in table \ref{tab:fltrs}.
Both the box and Gaussian filters are positive in physical space, which is important to 
guarantee physical realizability of filtered quantities \cite{Vremanetal94}.
On the other hand, the sharp spectral filter is not sign definite in x-space, 
which limits its utility in analyzing scale process in physical space.

Our results indicate an excellent agreement between baropycnal work and its nonlinear model. Using either the Gaussian or Box filters, the correlation coefficients are very high, $R_c>0.9$, for all the length scales we analyzed. The sharp spectral filter, on the other hand, yields poor agreement. This is not surprising since the sharp spectral filter can yield negative filtered densities \cite{Aluie13,ZhaoAluie18} and physically unrealizable subscale stresses \cite{Vremanetal94} due to its non-positivity in x-space. 

Figures \ref{fig:nl_model_box}, \ref{fig:nl_model_gauss}, and \ref{fig:nl_model_cutoff}, using the box, Gaussian, and sharp spectral filters, respectively,
plot $\Lambda(\bx)$ and $\Lambda_{\mModel}(\bx)$ along a line in the domain to show the typical agreement between the two quantities.
Also shown are the joint PDFs, which exhibit excellent linear agreement when using either the box or Gaussian kernels, but not the sharp spectral filter. Instantaneous visualizations in Figure \ref{fig:nl_model_slice} of $\Lambda(\bx)$ and $\Lambda_{\mModel}(\bx)$ are consistent with the excellent statistical agreement, showing an almost perfect pointwise correlation. We note that in our dilatationally forced flows, most of the contribution to $\Lambda_{\mModel}$ is from its straining component, $\Lambda_{SR}$, with a negligible contribution from $\Lambda_{BC}$ (see eq. \eqref{eq:lambda_decomp}). This is due to the shocks which contribute mostly to $\Lambda_{SR}$. In flows dominated by baroclinicity, such as in the Rayleigh-Taylor instability, a significant contribution to $\Lambda$ comes from $\Lambda_{BC}$, as will be shown in forthcoming work \cite{Zhaoetal19}.

\section{Summary}\lb{sec:Conclusion}
Past work \cite{Aluie11,Aluie13,EyinkDrivas18} has identified baropycnal work, $\Lambda$, as a process capable of transferring kinetic energy across scales in addition to deformation work, $\Pi$. This paper aimed at elucidating the physical mechanism by which $\Lambda$ operates.

Using scale-locality \cite{Aluie11} and a multiscale gradient expansion \cite{Eyink06a}, we derived a nonlinear model, $\Lambda_{\mModel}$, of baropycnal work. Using DNS, we showed excellent agreement between $\Lambda$ and $\Lambda_{\mModel}$ everywhere in space and at any time,  giving further empirical justification for our analysis of $\Lambda$ via its model $\Lambda_{\mModel}$.

We found that baropycnal work operates by the baroclinic generation of vorticity, and also by strain generation due to pressure and density gradients, both barotropic and baroclinic.
While the role of pressure and density gradients in generating vorticity is well recognized, their role in strain generation has been less emphasized in the literature.

As far as we know, this is the first direct demonstration of how baroclinicity
enters the kinetic energy budget, which arises naturally from our scale decomposition and the identification of $\Lambda$ as a scale-transfer mechanism.
Baroclinicity is often analyzed within the vorticity budget but its role in the energetics has never been obvious.
The need for a scale decomposition in order for $\Lambda$ and, as a result, baroclinic energy transfer, to appear in the kinetic energy budget is similar to the scale transfer term $\Pi$, which only appears in the budget after decomposing scales due to energy conservation. In the same vein, the appearance of baroclinicity in the vorticity equation can be interpreted as being a consequence of an effective scale decomposition performed by the curl operator $\grad\btimes$, which is a high-pass filter.
Our findings here support the argument in \cite{Aluie11,Aluie13} to separate $\Lambda$ from pressure-dilatation, $\OL{P}\grad\bdot\OL{\bu}$  in compressible LES, where the two terms are often lumped together in the form of $\OL{P}\grad\bdot\wt{\bu}$.

 In forthcoming work, we shall present further evidence of the excellent agreement between $\Lambda$ and $\Lambda_{\mModel}$ using low Mach number buoyancy driven flows with significant density variability \cite{Zhaoetal19}.

\vspace{6pt} 

 \section*{Acknowledgement}
This research was funded by DOE FES grant number DE-SC0014318. AL and HA were also supported DOE NNSA award DE-NA0003856. HA was also supported by NASA grant 80NSSC18K0772 and DOE grant DE-SC0019329. Computing time was provided by the National Energy Research Scientific Computing Center (NERSC) under Contract No. DE-AC02-05CH11231.
This report was prepared as an account of work sponsored by an agency of the U.S. Government. Neither the U.S. Government nor any agency thereof, nor any of their employees, makes any warranty, express or implied, or assumes any legal liability or responsibility for the accuracy, complete- ness, or usefulness of any information, apparatus, product, or process disclosed, or represents that its use would not infringe privately owned rights. Reference herein to any specific commercial product, process, or service by trade name, trademark, manufacturer, or otherwise does not necessarily constitute or imply its endorsement, recommendation, or favoring by the U.S. Government or any agency thereof. The views and opinions of authors expressed herein do not necessarily state or reflect those of the U.S. Government or any agency thereof.
\bibliographystyle{unsrt}
\bibliography{CompTurbulence}



\end{document}